\documentclass[conference]{IEEEtran}
\IEEEoverridecommandlockouts

\usepackage[table]{xcolor}
\usepackage{multirow}
\usepackage{subcaption}
\usepackage{orcidlink}
\usepackage{float}

\definecolor{MyDarkGreen}{rgb}{0,0.8,0.0}
\definecolor{MyDarkBlue}{rgb}{0,0,0.8}
\definecolor{MyDarkRed}{rgb}{0.6,0,0.0}

\def\BibTeX{{\rm B\kern-.05em{\sc i\kern-.025em b}\kern-.08em
    T\kern-.1667em\lower.7ex\hbox{E}\kern-.125emX}}

\AtBeginDocument{%
  \providecommand\BibTeX{{%
    \normalfont B\kern-0.5em{\scshape i\kern-0.25em b}\kern-0.8em\TeX}}}
    
\AtBeginDocument{%
 \providecommand\BibTeX{{%
 Bib\TeX}}}

\begin{document}

\title{The Bicameral Cache: a split cache for vector architectures\\

\thanks{This work has been supported by the Spanish Ministry of Science and Innovation under contract RED2022-134618-T, and together with the EU under contract PID2022-136454NB-C21 and grant TED2021-131176B-I00, as well as by the Spanish Ministry for Digital Transformation and Civil Service, together with the EU under contract TSI-069100-2023-0011.}
}

\author{
    \IEEEauthorblockN{
        Susana Rebolledo\orcidlink{0000-0002-1797-4202},
        Borja Perez\orcidlink{0000-0002-3695-2906}, 
        Jose Luis Bosque\orcidlink{0000-0002-7718-8449}}
    \IEEEauthorblockA{
        \textit{Dept. of Computer and Electronic Engineering} \\
        \textit{University of Cantabria}\\
        Santander, Spain \\
        \{rebolledos, perezpavonb, bosquejl\}@unican.es
    }
    \and
    \IEEEauthorblockN{Peter Hsu}
    \IEEEauthorblockA{
        \textit{Peter Hsu Consulting, Inc.} \\
        Barcelona, Spain and San Francisco, USA \\
         peterhsu3333@gmail.com
    }
  }

\maketitle

\begin{abstract}
The Bicameral Cache is a cache organization proposal for a vector architecture that segregates data according to their access type, distinguishing scalar from vector references. Its aim is to avoid both types of references from interfering in each other's data locality, with a special focus on prioritizing the performance on vector references. The proposed system incorporates an additional, non-polluting prefetching mechanism to help populate the long vector cache lines in advance to increase the hit rate by further exploiting the spatial locality on vector data. 
Its evaluation was conducted on the Cavatools simulator, comparing the performance to a standard conventional cache, over different typical vector benchmarks for several vector lengths.
The results proved the proposed cache speeds up performance on stride-1 vector benchmarks, while hardly impacting non-stride-1's. In addition, the prefetching feature consistently provided an additional value.

\end{abstract}


\section{Introduction}
One of the most common approaches to deal with the increasing computing demands of modern applications is vectorization. Based on exploiting data-level parallelism, this technique is not exclusive to supercomputers and HPC systems, but is also present in most commercial general-purpose processors \cite{vector-ref}, ranging from desktop to handheld devices. This has become even truer with the proliferation of new vector length agnostic (VLA) ISAs, such as ARM’s Scalable Vector Extensions (SVE) \cite{arm-sve} and RISC-V’s "V" (RVV) \cite{rvv}, which alleviate one of the challenges of vector processing: choosing the right vector length. In these extensions, the vector length used by the instructions is not fixed. It is specified at runtime, enabling the development of vector length agnostic (VLA) code that is compatible across different architectures, regardless of the actual physical vector length of the underlying hardware. This eases software development and portability and has had a significant influence on the renaissance of vector architectures as an option in the race for performance. 

 %Este párrafo no acaba de convencerme
In this scenario and considering its open-source nature, RVV is particularly relevant. Developed originally by the University of California, Berkeley in 2010, RISC-V has caught the attention of both academia and industry, due to its aim at standardization and focus on simplicity and extensibility. 

Despite potentially constituting a significant optimization, by operating over multiple data simultaneously, the effectiveness of vector instructions is inherently bound to the memory performance, which determines the speed at which such data becomes available \cite{musa_sato_cache_vp}. Reducing memory access latency or increasing bandwidth presents a critical challenge to bridge the Memory Wall gap \cite{mem-wall}. In addition, provided that access patterns and locality on scalar and vector references are noticeably different, incorporating dedicated memory solutions to exploit them can provide a significant gain, in both performance and energy consumption, in vector architectures. 

In this sense, one of the keys is adequately leveraging temporal and spatial locality, by using techniques such as the memory hierarchy, prefetching \cite{cache-prefetching} or bypassing \cite{bypassing_survey}. Temporal locality refers to the tendency to reference again, in the near future, elements that have been referenced recently. Spatial locality, on the other hand, is the tendency to reference, neighbouring elements to the one that has just been accessed. Scalar and vector references are fundamentally different regarding the kind of locality that they exhibit. The very semantics of vector instructions expresses that the same operation will be performed on several, often contiguous, data items. For this reason, this kind of instructions are good candidates for spatial locality. On the contrary, in a context in which vector instructions are used, scalar references often  mostly pose opportunities for temporal locality, as the spatial one is mainly exploited in a vector fashion.

This work addresses the semantic and behavioral differences between scalar and vector references, by proposing a cache segregated in two partitions with different organizations, that uses the mnemonics of the instructions themselves to split data based on access type. This proposed cache, called the Bicameral Cache, seeks to boost the performance of vector applications, by favoring the locality of vector data. Its dedicated partitions help to preserve the locality of both scalar and vector data, by using different cache line sizes tuned to each kind of access. This prevents scalar data from interfering with vector and vice versa. Additionally, the streaming nature of vector instructions is leveraged by using a memory-side prefetching scheme that opportunistically fills vector cache lines that belong to rows that are open in the memory controller. Experimental results show an average best-case speedup of 1.31x on stride-1 vector benchmarks, which rises up to 1.57x when using prefetching. Such values were obtained by a 22.43\% and 46.66\% improvements of the memory access time respectively.
Regarding non-stride-1 workloads, the Bicameral Cache achieves a best-case speedup of 1.11x when prefetching is used and no performance degradation overall when disabled.

The main contributions of this paper are:

\begin{itemize}
    \item It proposes a novel cache organization tuned to the inherent characteristics of vector and scalar references.
    \item It introduces a memory-side prefetching scheme that leverages the streaming nature of vector architectures. 
    \item It thoroughly evaluates the proposed memory hierarchy using a selection of stride-1 and non-stride-1 applications.
\end{itemize}

The rest of the paper is organized as follows: \autoref{sec:motivation} provides the background and motivation of this work, \autoref{sec:bc} thoroughly details the proposal, \autoref{sec:methodology} describes the methodology used in the evaluation,  \autoref{sec:results} presents and discusses the results, \autoref{sec:related_work} highlights the contributions and compares them with previous related work, and finally \autoref{sec:conclusion} summarises the results. 

\section{Background and motivation}
\label{sec:motivation}

This section introduces the fundamental memory and vector architecture concepts, including the operation of the cache, the memory controller and the additional pressure on memory generated by vector processing. It also presents the motivation for the proposal of the Bicameral Cache. 

\subsection{The memory subsystem}

The memory subsystem is one of the main factors of the performance of a computing system, as it plays a key role in hiding the latency of the access to data \cite{jacob}. It does so by leveraging locality: upon a memory reference, main memory is not accessed directly, but one or more caches of increasing size and decreasing speed are searched in sequence first, forming a hierarchy. This series of \textit{lookup} operations strives to keep data that is expected to be used close to the processor, in small memories that use faster technologies than the main memory. 

Considering its limited size, the adequate management of the contents of the cache is fundamental, regarding when data is brought into/pushed out of the cache and how updates are managed. The simplest way of pulling data into the cache is \textit{demand-fetching}: when the data associated to a lookup is not found, the cache line that contains it is brought into the cache from the next level in the hierarchy. More sophisticated schemes have also been explored, such as \textit{prefetching}, which is based on speculatively moving data that is expected to be used closer to the processor, even before it is referenced \cite{cache-prefetching}. 

When a cache cannot accommodate more data, a line needs to be selected for replacement. This decision is often made based on a heuristic, being \textit{Least-Recently-Used} (LRU) one of the most popular options. The update policy also has an impact on how replaced lines are managed. \textit{Write-through} policies update data both in the cache and in the next level of the memory hierarchy. \textit{Write-back} policies save bandwidth by only updating data in the cache. This implies that modified lines will be written into the backing store when they are replaced, increasing the cost of replacement. To alleviate this, a \textit{write buffer} \cite{write-buffer} is a common solution that removes writes from the critical path by introducing a buffer to temporarily store evicted lines that have been modified. These will be written back later, as time and resources permit, or may be reverted into the cache if referenced before the write takes place.

Locality also plays an important role in the access to main memory, which is usually implemented as a collection of DRAM banks that are split in rows and managed by a memory controller. Reads and writes in this kind of technologies are destructive, so every operation requires to back up the row that it involves, for later recovery. For this reason, each memory bank is equipped with a \textit{row-buffer} that stores the last accessed row. Rows are often long, so their transfer from/to the buffer adds an extra latency to the access to data, that can be avoided if the locality in the row buffer is properly leveraged. For this reason, accesses to an open row are often advantageous.

\subsection{Vector architectures and memory accesses}

Data-level parallelism (DLP), which consists in executing the same operation over different elements at the same time, is the basis of vector architectures. This approach means that, conceptually, a vector instruction acts as a set of scalar operations that are executed simultaneously on different data. The maximum vector length supported by a vector architecture (i.e. the size of its vector registers) determines the amount of data a single vector instruction can operate on. For instance, SVE supports a maximum vector length of 2048 bits, whereas the RVV allows registers of up to 65536 bits \cite{sve_arm} \cite{rvv}. The ideas proposed in this paper are evaluated using RVV, but they would apply to any other vector ISA.

Data parallel architectures are known for putting undue pressure on the memory hierarchy. Take for example the maximum vector length in RVV: if a vector memory instruction of an application operates on 64-bit elements, it means that it may generate as many as 1024 if the generated access pattern is not contiguous. Double the figure if the elements are 32-bit wide. If handled incorrectly, this amount of pressure may have several effects on the system, such as additional contention and an inadequate use of locality, both at the level of the cache and the row buffers in the memory controllers. The result is a degradation of the bandwidth of the system and performance overall. For this reason, memory hierarchy designs that are tuned to the specific characteristics of vector traffic are key to obtaining all the performance that these architectures are capable of delivering. In this sense, the semantics of vector instructions themselves pose a chance for proper prefetching, placement and locality schemes. 

\subsection{Motivation}

In a vectorized program, scalar and vector memory instructions coexist. These two types of instructions have very different semantics and behavior. Vector instructions access a series of memory addresses, generally contiguous, which greatly favours spatial locality. Scalar accesses, on the other hand, usually favor temporal locality, since spatial locality is better exploited with vector accesses. This difference means that the use of cache memories with different architectures, suitable for each type of locality, can provide a substantial advantage to the performance of these programs.

The spatial locality of vector operations is aided by a cache architecture with very long lines, divided into sectors that are managed independently, so as not to increase the miss penalty, as commonly used in GPUs \cite{amodt}. These long lines reduce the number of compulsory misses (i.e. data being brought into the cache for the first time). Additionally, this architecture allows the use of very simple, aggressive and non-speculative prefetching techniques, further reducing compulsory misses and exploiting the row buffer locality, allowing more main memory accesses to open rows, thus reducing their latency.

In addition, managing significantly fewer lines allows for a fully associative cache architecture, which is unfeasible when the number of cache lines is large, due to cost and latency overheads \cite{tagless}. %\cite{associativity}}. 
The fully associative cache eliminates conflict or collision misses, since any block can be placed in any line of the cache. These misses can degrade the performance of an application considerably, especially when replaced dirty blocks require to be updated in main memory.

On the contrary, scalar accesses are dominated by temporal locality, so they do not benefit from very long cache lines. For this access pattern, it is more profitable to have a large number of shorter lines. Hence, the cache can hold the necessary data for the time they are needed, to favor temporal locality. However, this implies that a set-associative architecture is needed to handle such number of lines.

Using a single set-associative cache for both types of references adds an additional problem. In set-associative caches, collisions occur, which force lines to be evicted from the cache. Given the mix of scalar and vector instructions, it is highly likely that scalar accesses will produce conflicts and force the replacement of vector data lines, incurring double penalty. First, the ability to leverage the spatial locality of the vector data may be lost, if the line is evicted before it is fully used. Second, it impairs the temporal locality of the scalar data, since they may not be reused before getting replaced.

For the reasons above, this paper proposes a Bicameral Cache, which leverages the locality of both types of instructions, eliminating the interference between them, and thus optimizing the performance of the memory hierarchy.

\section{The Bicameral Cache}
\label{sec:bc}

The proposed Bicameral Cache (BC) is a cache memory system for vector architectures composed of the combination of two data caches with different structures and geometries; the Scalar Cache and the Vector Cache. Its aim is to preserve the spatial data locality inherent to vector computation by preventing potential interference from the scalar traffic. The length of the Vector Cache lines, as well as the additional prefetching, or line-filling, capability, also enable a further exploitation of the spatial locality on vector data.

\begin{figure}[ht]
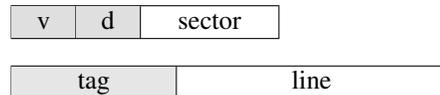

    \hspace*{4em}
    \begin{tabular}{|c|c|>{\centering\arraybackslash}m{4em}|}\hline
        \cellcolor{gray!25} {v } & \cellcolor{gray!25} {d } &  sector  \\
        \hline
    \end{tabular}

    \vspace{10pt}
    \hspace*{4em}
    \begin{tabular}{|>{\centering\arraybackslash}m{5em}|>{\centering\arraybackslash}m{9em}|}\hline
        \cellcolor{gray!20} {tag } & line  \\
        \hline
    \end{tabular}
\caption{Structural representation of lines and a sectors of the Bicameral Cache.}
\label{fig:line_sector_structure}
\end{figure}

Following a similar approach to the Sector Cache described in \cite{sector_cache}, the lines in both caches are sectorized, i.e., organized in sectors. Back in the day, sectoring first appeared as a solution to simplify the cache design according to the transistor logic, but it ultimately yielded performance improvements on multilevel cache designs, since it avoided over-fetching, therefore increasing both capacity and bandwidth utilization. This kind of design is commonly used in other throughput-focused architectures, such as GPUs \cite{amodt}.
In the Bicameral Cache, a sector is defined as the minimum data transfer unit between the main memory and the caches. Its  size is fixed to the length of the Scalar Cache lines. Vector Cache lines, on the contrary, are significantly longer, therefore composed of several sectors. Unlike former designs from \cite{sector_cache}, sectoring in the Bicameral Cache serves as a feature to exploit the spatial locality of vector data. \autoref{fig:line_sector_structure} depicts the structure of cache lines and sectors. While a line has, as usual, an associated tag to allow identification, each of its sectors present two bits to determine the state of their corresponding data; a valid bit (v) and a dirty bit (d). 

Both scalar and vector caches employ the Least Recently Used (LRU) replacement algorithm and a write-back policy. A representation of the BC structure is depicted in \autoref{fig:bc}.
%JLB: Esto me lo cargo para no poner números más que en la metodología. 
%\borrar{It is worth noting that, despite using specific sizes in \autoref{sec:results}, such as 64 B sectors and up-to-8-lines write buffers, the proposed cache is size-agnostic.}

\subsection{Scalar Cache}
The Scalar Cache (SC) stores data referenced by scalar memory instructions. It is organized using a set-associative mapping, with address fields identifying, from most to least significant, the tag, set and offset of the addressed data. 
%\borrar{, with 256 4-way sets in the evaluated configuration, which, considering 64 B lines (the same size as the sectors), sums up a total capacity of 64 KB.} 

Dirty lines need to be written back to main memory upon eviction. To reduce their impact on performance by preventing processor stalls, the Scalar Cache includes a write buffer (WB) in which evicted lines are temporally stored. Write buffer characteristics are explained on \autoref{sec:w-b-s}.

\begin{figure*}[h]
\centering
\begin{subfigure}{.43\textwidth}
\centering
    \begin{tabular}{|c|c|>{\centering\arraybackslash}m{2.5em}|}
        \hline
        \rowcolor{gray!40} Set & Line & Sector \\ \hline
        \multirow{4}{*}{\parbox[c]{0.5cm}{\centering 0}} & {\parbox[c]{0.5cm}{\centering 0}} &  \\  
        \cline{2-3}
        & {\parbox[c]{0.5cm}{\centering 1}} & \\
        \cline{2-3}
        & {\parbox[c]{0.5cm}{\centering 2}} &  \\ 
        \cline{2-3}
        & {\parbox[c]{0.5cm}{\centering 3}} &  \parbox[c]{1cm}{\centering\hfill} \\
        \hline
        \multirow{7}{*}{\parbox[c]{0.5cm}{\centering .\\ .\\ .}} & \multirow{7}{*}{\parbox[c]{0.5cm}{\centering .\\ .\\ .}} & \multirow{7}{*}{\parbox[c]{0.5cm}{\centering .\\ .\\ .}} \\  
        &  &  \\
        &  &  \\ 
        &  &  \\
        &  &  \\
        &  &  \\
        &  &  \\
        \cline{3-3}
        \hline
        \multirow{4}{*}{\parbox[c]{0.55cm}{\centering 255}} & {\parbox[c]{0.5cm}{\centering 0}} &  \\  
        \cline{2-3}
        & {\parbox[c]{0.5cm}{\centering 1}} & \\
        \cline{2-3}
        & {\parbox[c]{0.5cm}{\centering 2}} &  \\
        \cline{2-3}
        & {\parbox[c]{0.5cm}{\centering 3}} &  \parbox[c]{0.5cm}{\centering\hfill} \\
        \cline{3-3}
        \hline
    \end{tabular}
    \hspace{5pt}
    \begin{tabular}{|c|m{2.5em}|}
        \hline
        WB0 & \\
        \hline
        WB1 & \\
        \hline
        WB2 & \\
        \hline
        WB3 & \\
        \hline
        WB4 & \\
        \hline
        WB5 & \\
        \hline
        WB6 & \\
        \hline
        WB7 & \\
        \hline
    \end{tabular}
    \vspace{0.2em}
    \caption{Scalar Cache, on the left, and its write buffer, on the right.}
    \label{fig:sc}
    \vspace{1em}
\end{subfigure}
\hspace{1em}
\begin{subfigure}{.52\textwidth}
\centering
    \begin{tabular}{|c|*{5}{>{\centering\arraybackslash}m{2.5em}|}}
        \hline
        & \multicolumn{5}{c|}{\cellcolor{gray!40} Sector} \\
        \hline
        \cellcolor{gray!40} Line & 0 & 1 & $\cdots$ & 14 & 15 \\
        \hline
        0 & & & $\cdots$ & & \\
        \hline
        \multirow{4}{*}{\parbox[c]{0.5cm}{\centering .\\ .\\ .}} & \multicolumn{5}{c|}{\multirow{4}{*}{\parbox[c]{0.5cm}{\centering .\\ .\\ .}}} \\  
        & \multicolumn{5}{c|}{} \\
        & \multicolumn{5}{c|}{} \\
        & \multicolumn{5}{c|}{} \\
        \hline
        63 & & & $\cdots$ & &  \\
        \hline
        WB0 & \cellcolor{gray!20} & & $\cdots$ & \cellcolor{gray!20} &  \\
        \hline
        WB1 & & \cellcolor{gray!20} & $\cdots$ & \cellcolor{gray!20} & \\
        \hline
        WB2 & & & $\cdots$ & & \cellcolor{gray!20}  \\
        \hline
        WB3 & \cellcolor{gray!20} & \cellcolor{gray!20} & $\cdots$ & \cellcolor{gray!20} & \cellcolor{gray!20} \\
        \hline
        WB4 & \cellcolor{gray!20} &  & $\cdots$ &  & \\
        \hline
        WB5 & & \cellcolor{gray!20} & $\cdots$ &  & \cellcolor{gray!20} \\
        \hline
        WB6 & \cellcolor{gray!20} & \cellcolor{gray!20} & $\cdots$ & \cellcolor{gray!20} & \cellcolor{gray!20} \\
        \hline
        WB7 & & \cellcolor{gray!20} & $\cdots$ &  & \cellcolor{gray!20} \\
        \hline
    \end{tabular}
    \caption{Vector Cache. Sample shadowing helps identify modified sectors from write buffer lines awaiting write back to memory (i.e. valid and dirty).}
    \label{fig:vc}
\end{subfigure}
\caption{Structural representation of the Bicameral Cache.}
\label{fig:bc}
\end{figure*}

\autoref{fig:sc} shows a graphical representation of the SC structure, with 256 4-way sets, 64-B lines and an 8-line WB.

\subsection{Vector Cache}

In contrast, the Vector Cache (VC) stores data referenced by vector memory instructions, using longer lines that can fit several sectors. For a fixed cache size, this reduces the number of lines significantly, which allows the use of a fully associative cache, thus eliminating misses due to collision between lines located in the same set (conflict misses).

%\borrar{on a fully associative structure, which in the architecture modelled in \autoref{sec:results} holds 64 of 1024 lines, that fit 16 data sectors.}

This cache also implements a write-back policy and, considering the length of its lines, evictions have an even more significant impact on performance than in the Scalar Cache. However, cache lines this long would turn a conventional write buffer, into which modified evicted lines are migrated, into an expensive, slow and wasteful approach. For this reason, and as a design solution to make the most of the available capacity, the Vector Cache embeds its write buffer in the regular lines. When a modified line is evicted, it is simply flagged as WB, and will eventually be transferred to main memory or reverted to a regular cache line (more details in \autoref{sec:w-b-s}). This way, every WB line can be used as a regular cache line when available. Consequently, the effective cache capacity varies dynamically depending on the write buffer occupancy.

\autoref{fig:vc} presents the organization of the VC as a fully associative cache with 64 lines of 1024 bytes, each storing 16 sectors. From most to least significant positions, the address fields on the VC identify the tag, sector and offset, respectively.

\subsection{Scalar and vector cache exclusivity}

A particular memory address may be accessed by both vector and scalar memory instructions at different points of the execution of an application. If handled incorrectly, this situation may result in two independent (and potentially incoherent) copies of the same data existing in the vector and scalar caches. To prevent this, one key feature of the Bicameral Cache is the exclusivity between its two caches, which guarantees that a sector cannot be present in both of them at the same time. As a result, after a miss on the cache corresponding to the instruction type that referenced a given data (or \textit{native lookup}), the opposite cache shall be probed first (\textit{cross lookup}), before forwarding the request to main memory. On vector data exclusively, if data are found in the Scalar Cache (\textit{vector cross hit}), the corresponding sector is migrated from SC to VC. This one-sided migration policy aims at prioritising the presence of vector data in the Vector Cache, so as to favour the performance of the vector instructions. Thus, scalar memory instructions can reference vector data without disrupting the vector's continuity in the VC.

\subsection{Write-back strategy}
\label{sec:w-b-s}
%Esta entrada es un poco abrupta
%Sectors from write buffer lines can either be referenced again or written back to memory. 
The lifespan of a line in the write buffer (vector or scalar) extends until either (a) any of its valid sectors gets referenced again by a memory instruction or (b) it completes its write-back process. In both scenarios, the line leaves the write buffer.

In the first case, the line is fully restored to cache, either by replacing another existing line in the SC, or by simply reverting its write buffer status to a regular line in the VC. This line restoration property of the Bicameral Cache helps reduce access latency in the event of referencing freshly evicted data which is still held on the write buffer, due to the temporal locality of these data. 

In the second case, valid and dirty sectors of the line are sent, one at a time, to main memory. This process is triggered by the eviction of a cache line, with at least one valid and modified sector, that cannot join the write buffer due to full capacity. In this situation, the new referenced line cannot get its tag set in the cache until the eviction of the victim finishes. This requires the write-back requests for each of the valid and dirty sectors in the oldest line in the write buffer to be completed. Once done, the oldest line leaves the write buffer, making room for the victim line, which finally gets evicted from cache for the new referenced line to be placed. 

To avoid stalling the processor on this compulsory emptying, the Bicameral Cache implements an eager write buffer emptying scheme, which progressively and asynchronously writes back data into main memory. Right after requesting the referenced sector to memory, once both the native and cross lookups have missed, the write buffer occupancy is managed. If it reached a certain threshold, the transfer of the dirty sectors of the oldest WB line begins. 
Since the emptying time depends on the amount of valid and dirty sectors the line holds, this threshold was determined empirically for both caches, trying to find a trade-off between write buffer occupancy and bandwidth utilization to avoid undesired stalls. Hence, it was experimentally established to full capacity in SC, and to half plus one in the VC. 

%(i.e., 8 and 5 lines, respectively, in the analyzed scenario).
%\textcolor{red}{Estimating a timely effect to avoid undesired stalls, this threshold was experimentally established to full capacity in the SC and to half plus one in the VC, regarding the average amount of sectors in each line to write back. No se si esta frase está clara}

\subsection{Prefetching}

%BORJA: Si me he crecido con el "non-speculative", sentíos libres de cambiarlo
%BORJA: Creo que esta parte no queda muy clara
An additional feature of the proposal is memory-side prefetching of sectors of Vector Cache lines. This aims at filling vector lines in advance non-speculatively, to reduce the miss rate, preserve row buffer locality and improve the performance of vector operations. This prefetching mechanism is triggered when no idle memory bank has pending requests to serve, by scheduling a new access to the first available bank whose last operation was a read on a position before the end of a Vector Cache line (i.e., following sectors on the line are yet to be fetched and their corresponding row is still open). Trying to exploit row buffer locality, this access prefetches the sector that follows the one last read. It is important to note that this prefetch, in case of stride-1 operations, optimises the coverage and the accuracy, while it does not cause pollution, as all accessed data is used by the following vector operations. Furthermore, it does not evict data from the VC, as it is limited to filling existing lines with consecutive missing sectors.

% Cabe destacar que una única instrucción vectorial de acceso a memoria genera múltiples referencias, tantas como elementos tenga su vector, por lo que el número % depende del tamaño arquitectural establecido para este. Cada una de estas referencias realiza un acceso a memoria independiente generado en el mismo ciclo de manera simultánea. Para modelar este comportamiento, por simplicidad, se supone una única % referencia a la cache vectorial por cada sector.

%Creo que cabría o bien cerrar la sección o bien ir salpicándola con algo de justificación sobre porqué nuestra propuesta es buena y porqué obtiene buenos resultados.

\section{Methodology}
\label{sec:methodology}

The evaluation of the Bicameral Cache was performed using 64 B sectors and up-to-8-lines write buffers. For the Scalar Cache, a 256 4-way set configuration was used, which, considering 64 B lines (same size as the sectors), sums up a total capacity of 64 KB (as the example provided in \autoref{fig:sc}). For the Vector Cache, a fully associative cache of 1024 B per line (16 sectors), such as the one depicted in \autoref{fig:vc}, was considered. To match the SC capacity, 64 lines were needed.

To evaluate the proposal, Cavatools, an open-source RISC-V ISA simulator running on a x86 Linux system was used\footnote{Peter Hsu. Cavatools: https://github.com/phaa-eu/cavatools}. Its execution-driven scheme was modified to follow an event-driven approach that allows to model temporal constraints for each instruction. This implied incorporating a timing wheel, controlled by the system clock, on which every operation is scheduled to execute attending to their penalty cycles.

 The modelled architecture is a RISC-V single-core in-order vector processor, whose non-memory instruction latencies were estimated considering their complexity and the presumed characteristics of its pipeline according to the one described in \cite{latencies}. Latency values for non-memory vector instructions appear on \autoref{tab:lat_table}. Non-memory scalar instructions were assigned a 1 cycle latency.

\begin{table}[h]
\centering
\caption{Latencies for non-memory vector instructions.}
\begin{tabular}{|c|p{0.78\linewidth}|}
\hline
\rowcolor{gray!25}
\textbf{Latency} & \\
\rowcolor{gray!25}
\textbf{(cycles)} & \multirow{-2}{*}{\textbf{Vector instruction}}\\
\hline
\multirow{5}{*}{\parbox[c]{0.1cm}{\centering 4}} & vadd.vx, vadd.vv, vand.vx, vfabs.v, vfadd.vf, vfadd.vv, vfneg.v, vfsub.vf, vfsub.vv, vfcvt.f.x.v, vfcvt.x.f.v, vfmul.vf, vfmul.vv, vfmv.v.f, vmfle.vv, vmflt.vv, vmerge.vvm, vmseq.vv, vmul.vv, vmv.v.i, vmv.v.x, vmv1r.v, vmv.x.s, vor.vv, vsll.vi, vsrl.vi, vsub.vx \\
\hline
6 & vfmacc.vf, vfmacc.vv, vfmadd.vv, vfmul.vv \\
\hline
7 & vmin.vv \\
\hline
\multirow{2}{*}{\parbox[c]{0.1cm}{\centering 8}} & vredsum.vs, vfmax.vf, vfmax.vv, vfmin.vf, vslide1down.vx, vslide1up.vx \\
\hline
25 & vfdiv.vv, vfsqrt.v \\
\hline
\end{tabular}
\label{tab:lat_table}
\end{table}
 
The proposed Bicameral Cache was evaluated by comparing its performance to a single conventional scalar cache or white cache (WC), with same capacity of 128 KB. WC features and organization are similar to the Scalar Cache in BC: 4-way associativity, sector-size lines and a disjoint, 8-line write buffer. Furthermore, the effectiveness of prefetching was compared with an ideal version that provides maximum coverage by fetching all sectors in the VC line at once, with the same total latency of fetching the referenced one. In other words, this unrealistic behaviour would completely fill, without incurring any extra penalty, the whole vector line on the first compulsory miss that triggers a sector fetch from memory.

A simplified model of the main memory architecture was considered to simulate the behaviour of both BC and baseline, respectively, as part of the memory hierarchy. 
For the main memory, a 4 GB DRAM technology was chosen. Only RAS, CAS and precharge (PRE) operations were included in the modelling, whose penalties, extracted from \cite{bala_rajeev}, are reflected on \autoref{tab:ops}. The refresh and other complex operations, such as low power states, were not considered in the modelling of the memory controller. This is because a very high accuracy can be achieved by using a simple state machine with just two states, considering that what matters to a memory request is if the row to access is open or not \cite{memory_modelling_accuracy}. Each of the eight banks of this memory, distributed in 32768 rows and 256 columns, supports a single open row at a time. The address layout employed follows a standard Row-Bank-Column pattern.
To control the operations on the DRAM banks, a memory controller was also modelled. It is responsible for enqueuing lookup and write-back requests from the caches into each bank's queue, scheduling them for execution on a first-come-first-served basis. Communication between both caches and the memory controller has been modelled as a bi-directional 512-bit bus, the size of a sector. In addition, the memory controller also performs the prefetching of sectors for the Vector Cache lines.

\begin{table}[htbp]
    \centering
    \caption{Latencies for each memory operation.}
    \begin{tabular}{|l|c|}
        \hline
        \rowcolor{gray!25}
        \textbf{Operation} & \textbf{Latency (cycles)} \\
        \hline
        Native lookup (SC, VC, WC) & 1 \\
        Cross lookup (SC, VC) & 1 \\
        RAS & 28 \\
        CAS & 11 \\
        PRE & 11 \\
        \hline
    \end{tabular}
    \label{tab:ops}
\end{table}

Different types of well-known vector benchmarks were used in the evaluation: \texttt{axpy}, \texttt{blackscholes}, \texttt{jacobi-2d}, \texttt{lavaMD} and \texttt{pathfinder}, from the RiVEC suite \cite{ramirez}, and in-house implementations of matrix-matrix (\texttt{mm}), matrix-vector (\texttt{mv}) and sparse matrix-vector (\texttt{spmv}) multiplications. They present two different evaluation scenarios, according to their vector access pattern; stride-1 or non-stride-1. From the first category, \texttt{axpy} serves as an upper bound for the Bicameral Cache speedup over the white cache. The regions of interest in \texttt{blackscholes}, \texttt{pathfinder} and \texttt{spmv} were executed 100 times to obtain meaningful results, provided the huge time initialization takes in these benchmarks. \autoref{tab:benchmarks} lists the input size configurations used for each benchmark, as well as their access type group. In order to evaluate the proposal on a range of vector processors, several architectural vector lengths were used; 128, 256, 512, 1024, 2048 and 4096 bits.

\begin{figure*}[h]
    \centering
    \begin{subfigure}{\textwidth}
        \centering
        \includegraphics[width=0.95\textwidth]{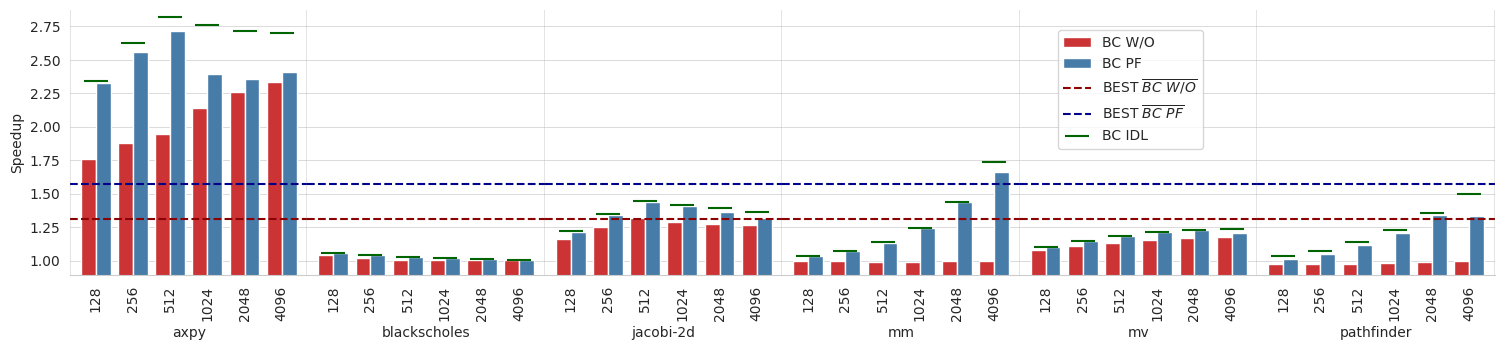}
        \caption{Stride-1 vector benchmarks.}
        \label{fig:sp-seq}
    \end{subfigure}
    \begin{subfigure}{\textwidth}
        \centering
        \includegraphics[width=0.95\textwidth]{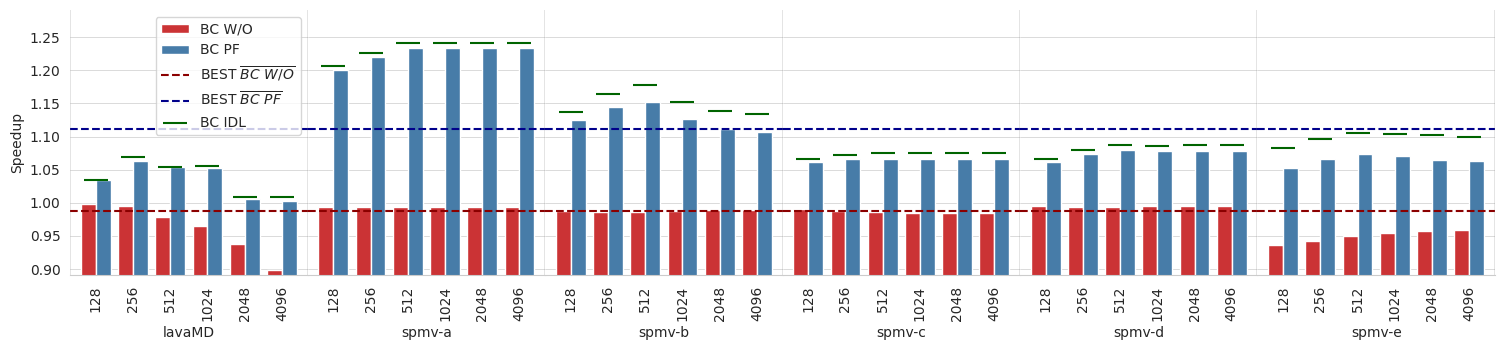}
        \caption{Non-stride-1 vector benchmarks.}
        \label{fig:sp-str}
    \end{subfigure}
    \caption{Speedup evaluation. BC: Bicameral Cache, W/O: without prefetching, PF: with prefetching, IDL: with ideal prefetching.}
    \label{fig:speedup}
\end{figure*}
\begin{figure*}[h!]
    \centering
    \begin{subfigure}{\textwidth}
        \centering
        \includegraphics[width=0.95\textwidth]{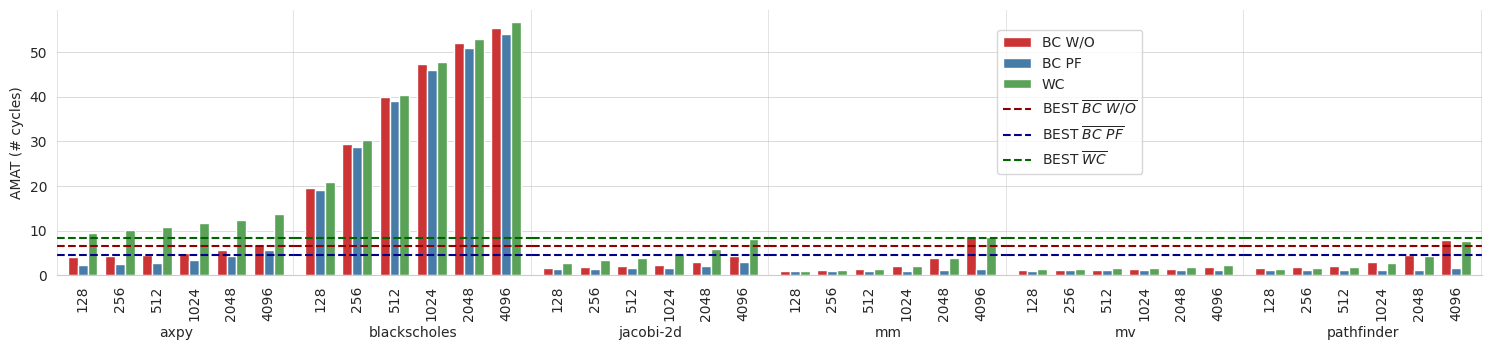}
        \caption{Stride-1 vector benchmarks.}
        \label{fig:amat-seq}
    \end{subfigure}
    \begin{subfigure}{\textwidth}
        \centering
        \includegraphics[width=0.95\textwidth]{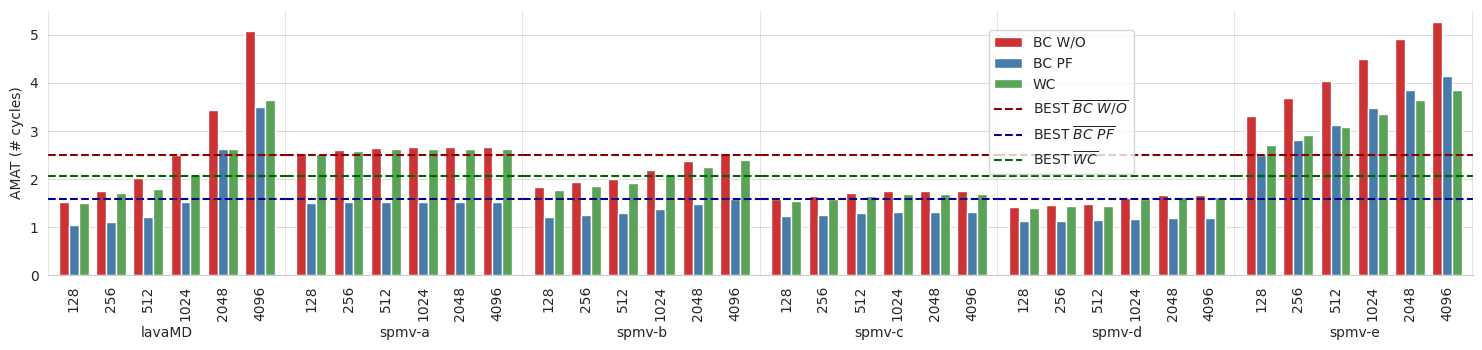}
        \caption{Non-stride-1 vector benchmarks.}
        \label{fig:amat-str}
    \end{subfigure}
    \caption{Average Memory Access Time. BC: Bicameral Cache, W/O: without prefetching, PF: with prefetching, IDL: with ideal prefetching.}
    \label{fig:amat}
\end{figure*}

\begin{table}[htbp]
\caption{Benchmark configurations used in the evaluation.}
\begin{center}
\begin{tabular}{l*{2}{c}}
\hline
Benchmark & Input size  & Stride \\
\hline
\texttt{axpy} & 2048 KB  & 1\\
\texttt{blackscholes} & 131072 opt. & 1  \\
\texttt{jacobi-2d} & 256 el. & 1 \\
\texttt{pathfinder} & 4096x4096 el.  & 1\\
\texttt{mm} & 256x256 el.   & 1\\
\texttt{mv} & 4096x4096 el.   & 1\\
\texttt{lavaMD} & 8192 par. & $\neq$ 1 \\
\texttt{spmv-a} & ncvxqp9 (ID=1243) \cite{smc} & $\neq$ 1 \\
\texttt{spmv-b} & rajat15 (ID=1316) \cite{smc}& $\neq$ 1 \\
\texttt{spmv-c} & rail\_79841 (ID=1445) \cite{smc} & $\neq$ 1 \\
\texttt{spmv-d} & benzene (ID=1348) \cite{smc} & $\neq$ 1 \\
\texttt{spmv-e} & cond-mat-2005 (ID=2395) \cite{smc} & $\neq$ 1 \\
\hline
\end{tabular}
\end{center}

\footnotesize{opt.: options, el.: elements, par.: particles.}
\footnotesize{Matrices used in the \texttt{spmv} evaluations where extracted from \cite{smc}.}
\label{tab:benchmarks}
\end{table}

\section{Evaluation}
\label{sec:results}

The goal of the experimental evaluation is to evaluate the performance gain obtained by the Bicameral Cache over the conventional cache described above. To this end, two metrics have been analysed: the speedup, presented in \autoref{fig:speedup}, and the average memory access time (AMAT), shown in \autoref{fig:amat}, both with and without prefetching.

%JLB: Párrafo resumen de las medias de los resultados. 
The main conclusion that can be highlighted from these results is that the Bicameral Cache with prefetching improves the baseline in all cases analysed. On average, on stride-1 vector benchmarks, it achieves an improvement of roughly 1.31x (without prefetching) and 1.57x (with prefetching), when considering the mean of the best performing vector lengths, compared to the conventional cache, as depicted in \autoref{fig:sp-seq}. These values derive from a significant decrease in the average memory access time of the best case, which drops from 8.47 cycles on a conventional cache to 6.57 and 4.52 cycles, respectively, as shown in \autoref{fig:amat-seq}.

On non-stride-1 vector benchmarks, the Bicameral Cache with prefetching achieves a best-case average gain of 11\%, with peak values of up to 1.23x, illustrated in \autoref{fig:sp-str}. This corresponds to roughly a 0.47 cycle decrease in the AMAT values, as shown in \autoref{fig:amat-str}. When prefetching is not used, BC cannot improve the best-case baseline, with degradation values of about 1\% in speedup and 0.29 cycles in AMAT.

%JLB: Tamaño del vector.
With regard to the vector length, the results show that there is an optimal value for each benchmark, which must be determined empirically. Only \texttt{mm} seems to be able to take advantage of longer vectors (this behaviour is analysed in more detail later). In many cases, the performance variations using different vector sizes are quite significant, so selecting the best length for each application is critical.

%JLB: Falta comentar axpy, blackscholes y mv. 

Analysing the benchmarks individually, it can be seen that in the case of stride-1, \texttt{axpy},  \texttt{blackscholes}, \texttt{jacobi-2d} and \texttt{mv} improvements appear for both BC versions -- with and without prefetch -- on any vector length. For \texttt{mm} and \texttt{pathfinder}, the performance of the proposal slightly drops, in absence of prefetching, up to 3\% in the worst case, yet if enabled, it significantly outperforms the baseline, reaching even higher values, generally, as vector length increases (\autoref{fig:sp-str}). This behaviour is also reflected in the AMAT values from \autoref{fig:amat-str}, which are not reduced in the aforementioned cases without prefetching. 

In non-stride-1 benchmarks, the BC without prefetching does not outperform the WC. This is reflected in the AMATs, in which the time of the BC without prefetching is always higher or equal to that of the white cache. When including prefetching, \texttt{lava} achieves a moderate improvement (6\% at best), which is strongly dependent on the vector length. In \texttt{spmv}, the vector size is found to have a much lower impact on the performance than the sparsity of the input matrix, with values ranging from a gain of 23\% to 7\%.

Overall, prefetching not only does not degrade performance in any of the experiments, but it consistently improves the BC results to values very close -- even equal -- to the ideal upper bound. It can thus be considered a recommended optimization.

\begin{figure*}[htbp]
    \begin{center}
    \begin{subfigure}{0.32\textwidth}
        \includegraphics[width=\textwidth]{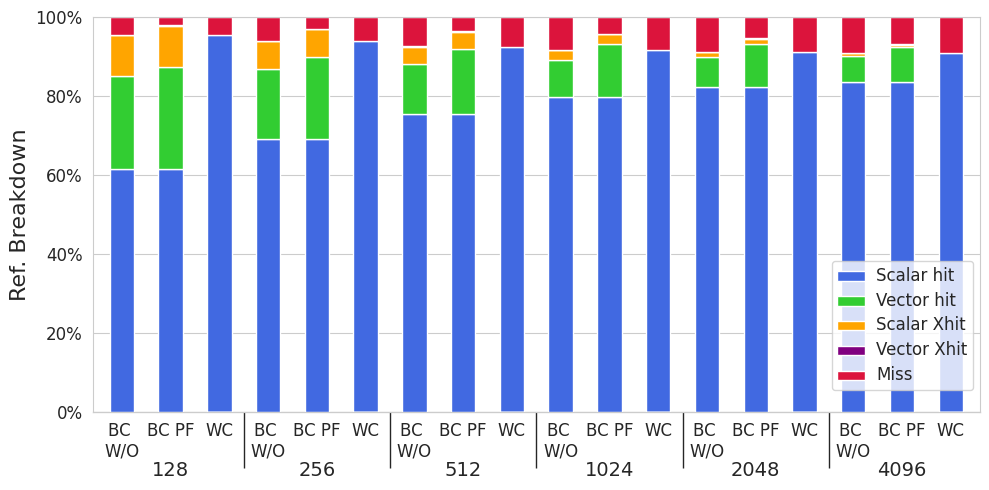}
        \caption{Reference breakdown: \texttt{jacobi-2d}.}
        \label{fig:ref_bd_jac}
    \end{subfigure}
    \begin{subfigure}{0.32\textwidth}
        \includegraphics[width=\textwidth]{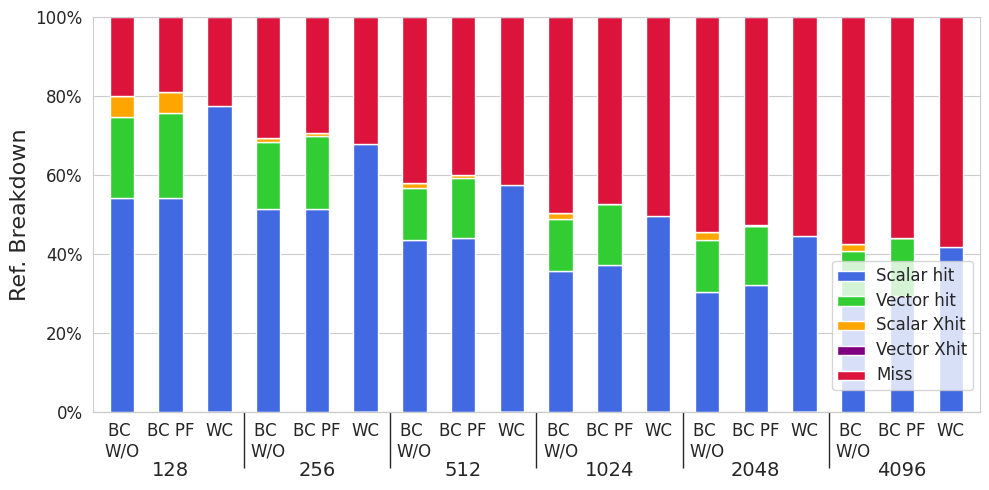}
        \caption{Reference breakdown: \texttt{blackscholes}.}
        \label{fig:ref_bd_bk}
    \end{subfigure}
    \begin{subfigure}{0.32\textwidth}
        \includegraphics[width=\textwidth]{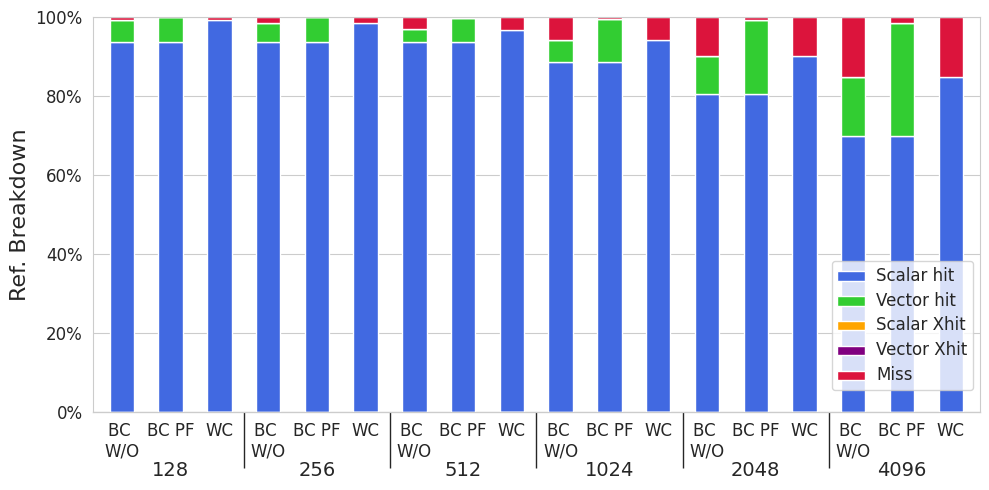}
        \caption{Reference breakdown: \texttt{mm}.}
        \label{fig:ref_bd_mm}
    \end{subfigure}
    \begin{subfigure}{0.32\textwidth}
        \includegraphics[width=\textwidth]{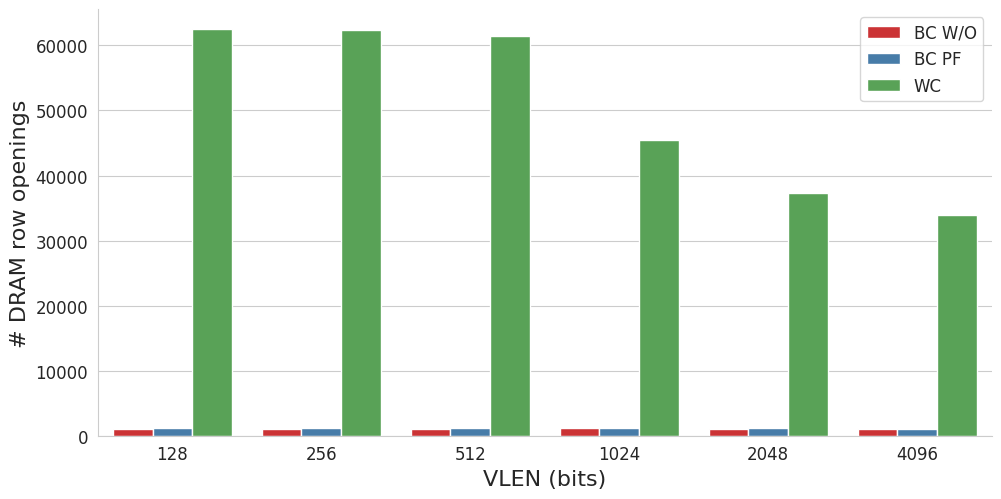}
        \caption{DRAM row openings: \texttt{jacobi-2d}.}
        \label{fig:op_jac}
    \end{subfigure}
     \begin{subfigure}{0.32\textwidth}
        \includegraphics[width=\textwidth]{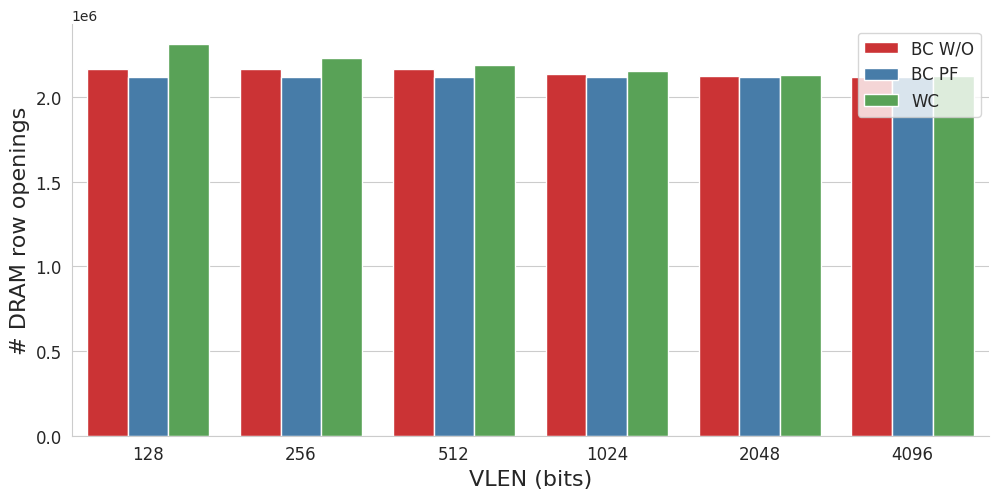}
        \caption{DRAM row openings: \texttt{blackscholes}.}
        \label{fig:op_bk}
    \end{subfigure}
     \begin{subfigure}{0.32\textwidth}
        \includegraphics[width=\textwidth]{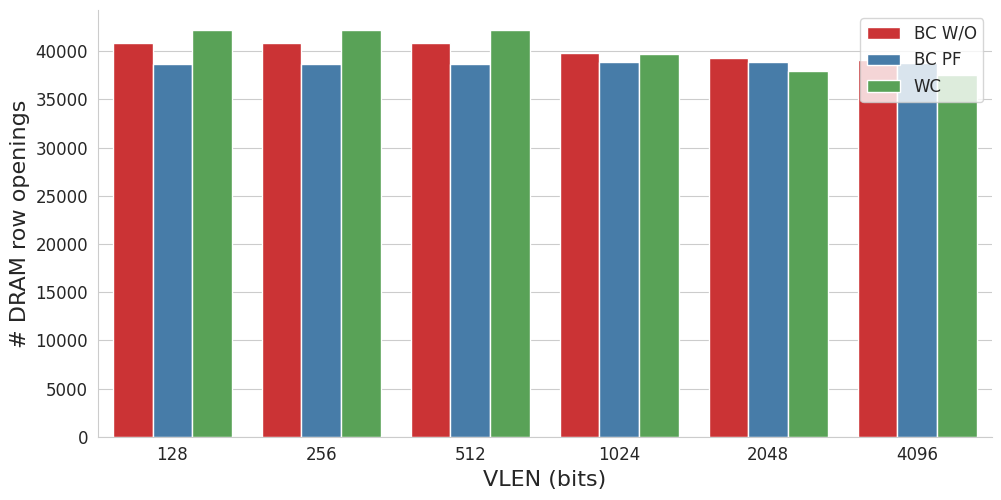}
        \caption{DRAM row openings: \texttt{mm}.}
        \label{fig:op_mm}
    \end{subfigure}
    \end{center}
    \centering
    \caption{Detailed analysis. BC: Bicameral Cache, W/O: without prefetching, PF: with prefetching, IDL: with ideal prefetching.}
    \label{fig:analysis}
\end{figure*}

In order to better understand the source of such performance improvements, a detailed analysis of \texttt{jacobi-2d}, \texttt{blackscholes} and \texttt{mm} is presented in \autoref{fig:analysis}. Their reference breakdown is shown in figures \ref{fig:ref_bd_jac}, \ref{fig:ref_bd_bk} and \ref{fig:ref_bd_mm}, while figures  \ref{fig:op_jac},  \ref{fig:op_bk} and  \ref{fig:op_mm} present the number of DRAM row openings for each benchmark.

First, the reference breakdown of \texttt{jacobi-2d} in \autoref{fig:ref_bd_jac} helps clarify the prefetching effect; by filling vector lines in advance, the miss rates are significantly lowered, since more data is found on cache when referenced, which explains the observed average latency reduction on memory accesses. Nevertheless, speedups were not simply achieved when enabling this prefetching. \autoref{fig:op_jac} shows how, as expected, simply by reorganizing data in cache, the memory overhead drastically decreases. Specifically on \texttt{jacobi-2d}, the number of DRAM row openings when using the Bicameral Cache drops at least a 96\%. Row openings conform the worst-case scenario when handling a memory request on a DRAM bank, since they require the longest time on closing the previously open row (PRE), activating the new row (RAS) and reading its column (CAS). On a similar miss rate context, as it is the case for the Bicameral Cache without prefetching and the white cache, such row openings increase is due to the write-back requests. Since all data are treated the same, the write buffer in the conventional cache does not distinguish its lines according to their reference type. As a result, consecutive write-backs can provoke bank collisions on main memory, as they can be one of each type (scalar after vector, and vice versa). Therefore, both memory overhead and bandwidth utilization increase with the amount of row openings which, even though most of the times the processor does not stall, as it seems to be the case, due to the preventive write-back strategy, is enough to contribute in the performance improvement of the non-prefetching version of the BC over the baseline. \texttt{axpy}, \texttt{mv} and \texttt{pathfinder} have a very similar behavior to \texttt{jacobi-2d}. However, prefetching reduces significantly more the miss rate in \texttt{pathfinder}, having a greater influence in the observed improvements.

The low improvement obtained in \texttt{blackscholes} is explained by Figures \ref{fig:ref_bd_bk} and \ref{fig:op_bk}. The miss rate is very high, exceeding 50\% for large vectors. Prefetching in BC reduces this high miss rate, but not significantly. In addition, the number of row openings in DRAM remains almost constant, both with and without prefetch.

Finally, in \texttt{mm}, \autoref{fig:op_mm} shows that the number of DRAM openings, which is two orders of magnitude lower than in \texttt{blackscholes}, is also very slightly reduced, even increased for the longest vector sizes. However, in this case, \autoref{fig:ref_bd_mm} shows that BC does achieve a substantial improvement in the miss rate using prefetch, especially for large vector sizes.

\section{Related work}
\label{sec:related_work}

%Esto hay que contarlo con cuidado para que no nos den un incremental work
%The Bicameral Cache concept employs a combination of existing design ideas and innovative strategies to achieve a unified joint solution. 

%%BORJA: Esto está muy bien, pero es un poco largo. Creo que tenemos también que añadir alguna referencia más a otros trabajos. La semana que viene puedo buscar un poco.

Hardware data prefetching is a common mechanism to improve performance by reducing miss rates. This method requires supporting hardware components, and has been widely covered in several prior works through different techniques, including temporal \cite{domino} \cite{temporal_data_pf} \cite{hao_temp}, spatial \cite{bingo} \cite{bouquet} and based on deltas \cite{berti} \cite{complex_delta_pf} and predictors \cite{prediction_pf} approaches, among others. However, efforts to develop specific prefetching mechanisms for vector architectures have been more limited. Some examples present specific prefetching schemes for multiprocessor vector caches \cite{fu&patel} or complement existing ones for vector processors by exploiting code semantics \cite{revela}. The prefetching presented in this work is a simple, yet effective, mechanism especially designed for a vector architecture. Its simplicity and non-polluting nature avoid frequent negative effects such as hardware/area/power overheads and cache pollution.

While some previous works have focused on developing and evaluating cache systems for vector architectures \cite{vru} \cite{media_oriented_vp} \cite{noel-v} \cite{shared_cache_vp}, the base idea of splitting data in cache in accordance to their type of locality was introduced in \cite{dual_data_cache}, where a prediction table was used to determine which data was placed on which subcache. In contrast, the Bicameral Cache simply sorts data based on their reference type, assuming only temporal locality on scalar data and both spatial and a certain degree of temporal locality on vector data. The design in \cite{dual_data_cache}, referred to as the Dual Data Cache, also allowed different versions of the same data element to reside in both caches simultaneously. Apart from requiring additional steps to preserve coherence by updating the old copy once the valid one gets evicted, caching duplicated values leads to inefficient capacity management, since it wastes valuable resources. The Bicameral Cache, on the other hand, adopts a mutual exclusivity policy with a vector migration feature that helps to further prioritize the vector's continuity while also allowing scalar instructions to work with vector data without compromising their spatial locality. Another substantial difference with the work in \cite{dual_data_cache} is the write buffer management. Despite mentioning a write-back strategy, no specific implementation, other than the updating of copies in cache, is detailed in \cite{dual_data_cache}. Instead, the BC integrates two distinct write buffer concepts, tailored to the specific characteristics of each cache; a disjoint, conventional structure for the Scalar Cache and an embedded, dynamic one in the vector part.  Finally, this proposal further differentiates from the Dual Data Cache on both prefetching scheme and cache organization. Whereas the previous design assumes two direct-mapping subcaches with a simple next-line prefetching, which can pollute the cache, the proposed model considers the needs of each structure, distinguishing a common set-associative organization on the scalar partition and a fully associative structure on the vector one. Following a similar approach to the Sector Cache in \cite{sector_cache}, cache lines in BC are sectorized with the purpose of dedicating vector lines to store consecutive elements of the same vector. This enables the prefetching technique to serve as a relevant vector filling aid which, inspired by similar approaches in \cite{fu&patel} and \cite{batten}, simply brings following sectors to existing vector lines, thereby enhancing performance by reducing the miss rate while also avoiding undesired cache pollution and replacements.

\section{Conclusion}
\label{sec:conclusion}

This paper presents the design and evaluation of a proposed cache memory for a vector processor which splits scalar and vector references into two partitions with different characteristics.
The proposal, called Bicameral Cache, is specifically aimed at improving the performance of vector applications, not only by suppressing the possible interference caused by scalar instructions on vector data, but also by arranging vector elements consecutively to guarantee their continuity. In addition, a prefetching, or data prefilling, option that populates the vector cache lines is included to optimise performance by further exploiting the spatial locality in the vector references. 
The Cavatools simulator, which supports the RISC-V vector extension, was used for the implementation. To evaluate the proposal against a conventional cache, simulations of several vector benchmarks with a variety of architectural vector lengths have been conducted. 
The results indicate that the proposed cache is advantageous for stride-1 vector benchmarks, achieving an average best-case speedup of 1.31x, which grows up to 1.57x if enabling the prefetching.
Furthermore, on non-stride-1 workloads, the performance is barely degraded with the basic configuration, yet it improves on an 11\% on the average best-case scenario with prefetching.
In conclusion, these results represent a notable enhancement as they are achieved without additional hardware or an increase in the cache size; just by restructuring and leveraging the available resources. Finally, this prefetching feature is highly recommended, as it consistently provides an added value.

\bibliographystyle{IEEEtran}
\bibliography{IEEEabrv,bib}

\end{document}